   \documentclass[aps,twocolumn,prb,showpacs]{revtex4}

\usepackage{graphicx}
\usepackage{latexsym}
\usepackage{amsmath}
\usepackage[colorlinks,
bookmarks=false, citecolor=blue, linkcolor=red, urlcolor=blue]{hyperref}

\begin{document}
\title{Finite-temperature phase transitions in the quantum fully frustrated
Ising models} 
\author{S.\,E. Korshunov}
\affiliation{L.\,D. Landau Institute for Theoretical Physics RAS, 142432
Chernogolovka, Russia}
\date{July, 2012}

\begin{abstract}
The quantum antiferromagnetic spin-1/2 Ising model on a triangular lattice
and analogous fully frustrated Ising model on a square lattice with
quantum fluctuations induced by the application of the transverse magnetic
field are studied at finite temperatures by constructing an exact mapping
onto a purely classical model with a more complex interaction. It is shown
that in weak fields the temperatures of the phase transitions separating
the critical phase from the ordered and disordered phases in both  models
are proportional to the magnitude of the field.

\end{abstract}

\pacs{05.50.+q, 75.10.-w, 71.10.Jm, 75.50.Ee} 

\maketitle

\section{Introduction}

In recent years the interest to quantum frustrated magnetic systems is
constantly growing due to the permanent appearance of new materials
belonging to this class (see Ref. \onlinecite{review} for a recent
review). In many cases, an interplay between fluctuations (quantum and
thermal) and a macroscopic degeneracy of the classical ground states makes
understanding the properties of such systems a nontrivial task.

The best known example of a frustrated magnetic system is the
antiferromagnetic Ising model on a triangular lattice, in which each
triangular plaquette has to contain at least one frustrated bond with
higher than minimal energy. It is well known from the exact solution
\cite{Wan} that this classical model is disordered at any positive
temperature $T>0$, whereas at $T=0$ it is characterized by  a finite
residual entropy and an algebraic decay of correlations. \cite{Steph} A
quantum analog of this model can be defined by the Hamiltonian,
\begin{equation}                                                 \label{H}
\hat{H}=J\sum_{\left( jj'\right)}{{}\hat\sigma}_j^z{{}\hat\sigma}_{j'}^z
 -\Gamma\sum_{j}{{{}\hat \sigma}_j^x}\,,
\end{equation}
where $J>0$ is the coupling constant of the nearest-neighbor interaction,
$\Gamma>0$ is proportional to the magnitude of the transverse magnetic
field inducing quantum fluctuations, ${}\hat \sigma^x$ and ${}\hat
\sigma^z$ are the Pauli matrices, and the summation in the first term is
performed over all pairs of nearest neighbors $\left(jj'\right)$ on a
triangular lattice. The classical antiferromagnetic Ising model is
recovered in the case $\Gamma=0$.

Besides being applicable for the description of an easy-axis
antiferromagnet with spin $1/2$ and a triangular lattice, the quantum
antiferromagnetic Ising model defined by Eq. (\ref{H}) is also of interest
as a representative of a wider class of the fully frustrated
transverse-field Ising models, in which coupling constants can have both
signs \makebox{($J_{jj'}=\pm J$)}, but on each plaquette of the lattice
the number of the antiferromagnetic bonds (with $J_{jj'} = J>0$) is odd.
\cite{V77} The quantum fully frustrated Ising models on various lattices
have been investigated \cite{MSCh,MS,Mila,Coletta,WCKM,HHMR} mostly in
view of their relation to the quantum dimer models \cite{MR} with
vanishing potential energy term. A method for constructing more general
models implementing a continuous crossover between a frustrated Ising
model and a dimer model (on the dual lattice) whose Hamiltonian includes
both the kinetic and potential terms has been proposed recently in Ref.
\onlinecite{IvK}.

The first investigation of the finite-temperature phase diagram of the
quantum antiferromagnetic Ising model on a triangular lattice was
undertaken by Isakov and Moessner. \cite{IM} Relying on the analogy with
the classical antiferromagnetic Ising model on a layered triangular
lattice \cite{BMBGS} in a system with a finite size in the direction
perpendicular to the layers they argued that at
\makebox{$0<\Gamma<\Gamma_c$} (where \makebox{$\Gamma_c\propto J$} is the
critical value of $\Gamma$ at zero temperature) this model has to have two
finite-temperature phase transitions of the
Berezinskii\--Kosterlitz\--Thouless type with a critical phase between
them (like in the classical six-state clock model) and confirmed these
conclusions with the help of numerical simulations. Soon after that, Jiang
and Emig \cite{JE} proposed a derivation (based on a renormalization group
analysis) demonstrating that the temperatures of these two transitions
$T^c_1$ and $T^c_2$ should behave themselves as
\begin{equation}                                 \label{Tc}
T^c_{1,2}\propto\Gamma\ln^\nu(\Gamma_c/\Gamma)\,,
\end{equation}
where $\nu$ is the critical exponent of the three-dimensional XY model
and \makebox{$T^c_2/T^c_1=9/4$}.

However, the prediction of Ref. \onlinecite{JE} was based on the renormalization-group
analysis assuming the possibility of describing the evolution of the system
in Euclidean time in the framework of the continuous approximation.
This approach requires the size of the system in the time direction $\beta=1/T$
to be much larger than the typical time between spin flips (inversely
proportional to $\Gamma$), which at temperatures given by Eq. (\ref{Tc})
is fulfilled only for \makebox{$\Gamma_c-\Gamma\ll \Gamma_c$}. 

The main aim of the present work is to study the behavior of $T^c_1$
and $T^c_2$ in the quantum antiferromagnetic Ising model on a triangular
lattice
at $\Gamma\ll J$, where the continuous approximation is not applicable.
Our result consists in finding that in this range of parameters
both transition temperatures
are just proportional to $\Gamma$ and do not diverge with the increase of
$J$ as predicted by Eq. (\ref{Tc}). We also demonstrate that in the fully
frustrated transverse-field Ising model on a square lattice the situation
is the basically the same: the ordered and disordered phases are separated
by the critical phase with both transition temperatures being proportional
to the magnitude of the field in weak fields. Up to now, the investigation
of this model has been focused on its zero-temperature properties.
\cite{MSCh,MS,WCKM}

\section{Antiferromagnetic model on a triangular lattice}
\subsection{The case of the infinite coupling constant}

We start by considering the quantum antiferromagnetic Ising model on a
triangular lattice in the special case of the infinite coupling constant,
$J=\infty$.
In such a case, the Hilbert space of the model is restricted to the states
with only one frustrated bond in each triangular plaquette (that is, the
ground states of the classical antiferromagnetic Ising model labelled
below by index ${a}$) and their linear combinations. The
finite-temperature partition function can be then written as
\begin{equation}                                      \label{Z}
Z=\sum_{{a}}^{}W_{a}\end{equation}
with
\begin{equation}                                       \label{W}
W_{a} =\langle{a}|\exp(-\beta \hat H)|{a}\rangle
\end{equation}
and $\beta \equiv 1/T$.

In the infinite-temperature limit ($\beta\to 0$) one gets
$W_{a} \to 1$, that is, $Z$ is reduced to the {\em zero-temperature}
partition function of the classical antiferromagnetic Ising model.
At $\beta>0$ weights $W_{a}$ become dependent on the structure of the state ${a}$.
A convenient way for describing this dependence consists in introducing
an effective classical Hamiltonian
${\cal H}_{\rm eff}(a)$ defined by the relation
\begin{equation}
W_{a}\equiv \exp[-{\cal H}_{\rm eff}({a})]\,.
\end{equation}
At $\beta\Gamma\ll 1$ the spin flips induced by the second term in Eq.
(\ref{H}) are rare which allows one to calculate the statistical weights
$W_a$ corresponding to different states perturbatively. In such a way, one
obtains that in the lowest order in $\beta\Gamma$, ${\cal H}_{\rm eff}(a)$
is just proportional to $M_{\rm }({a})$, the number of spins in state
${a}$ each of which can be flipped without taking the system out of its
Hilbert space,
\begin{equation}                                       \label{Heff}
{\cal H}_{\rm eff}({a})=-\frac{1}{2}(\beta \Gamma)^2M_{\rm }({a})\,.
\end{equation}
The terms of the fourth and higher orders in $\beta$ depend on the
structure of the state $a$ in a more complex way, but at $\beta\Gamma\ll
1$ the main role is played by the terms included into Eq. (\ref{Heff}).

It is well known that the set of the ground states of the classical
antiferromagnetic Ising model on a triangular lattice can be mapped onto
the states of a solid-on-solid (SOS) model describing $[111]$ facet of a
cubic crystal. \cite{BH} In terms of the SOS representation, each Ising
spin $\sigma_j=\pm 1$ is replaced by integer height variable $h_j$ in such
a way that the relation
\begin{equation}                                               \label{h-h'}
    h_{j'}-h_j=-\frac{1+3\sigma_j\sigma_{j'}}{2}=\left\{
    \begin{array}{l} +1\\
    -2\end{array}\right.
\end{equation}
is satisfied for every pair of neighboring sites on the triangular
lattice. Eq. (\ref{h-h'}) assumes that one of the three basic vectors of
the triangular lattice (whose sum is equal to zero) is directed from site
$j$ to site $j'$ and not from $j'$ to $j$ (which would correspond to the
opposite sign of $h_{j'}-h_j$).

When each triangular plaquette contains only one frustrated bond (with
$\sigma_j\sigma_{j'}=+1$), Eqs. (\ref{h-h'}) unambiguously define the
values of all integer variables $h_j$ as soon as the value of $h_j$ is
chosen for one of the sites. On the other hand, any state of the SOS model
with $h_{j'}-h_j=+1,-2$ corresponds to some ground state of the
antiferromagnetic Ising model.

It follows from the known properties of the classical antiferromagnetic
Ising model on a triangular lattice that when all allowed states of such
an SOS model enter the partition function with the same weight, this model
is in the rough phase, in which correlations of heights  diverge
logarithmically, \cite{NHB}
\begin{equation}                                        \label{g}
g_{jk}=\langle(h_j-h_k)^2\rangle \approx \frac{K}{2\pi^2} \log r_{jk}\,,
\end{equation}
where $r_{jk}$ is distance between sites $j$ and $k$
and \makebox{$K=K_0=18$}.

The same surface representation can be used for interpreting the quantum
spin model (\ref{H}) with $J=\infty$ as a quantum SOS model in which the
amplitude of height jumps, $h_j\to h_j\pm 3$ is given by $\Gamma$.
Alternatively, the use of Eq. (\ref{W}) allows one to speak about the
classical SOS model in which now the weights corresponding to different
configurations of the surface are no longer equal to each other and depend
on the configuration.
Since in terms of the height representation flippable spins correspond to
points where the surface has no local slope, in terms of the SOS
representation functional $M(a)$ is a measure of the flatness of the
surface. Then it is clear from the negative sign in Eq. (\ref{Heff}), that
the interaction described by this equation suppresses the fluctuations of
the surface.

It has to be emphasized that the classical SOS model defined by the
Hamiltonian ${\cal H}_{\rm eff}(a)$ is exactly equivalent to the original
quantum spin model.
The price one has to pay for the reduction of a quantum model at a finite
temperature to a purely classical one is that it is impossible to write
down the explicit form of ${\cal H}_{\rm eff}(a)$, Eq. (\ref{Heff}) being
applicable only at $\beta\Gamma\ll 1$. However, a number of properties of
the classical SOS model defined by ${\cal H}_{\rm eff}(a)$ can be
discussed without knowing the exact form of ${\cal H}_{\rm eff}(a)$.

In particular, it is clear that the decrease of $T$ (increase of
$\beta\equiv 1/T$) leads to the suppression of the factor $K$ in the
correlation function (\ref{g}) and finally has to induce a phase
transition into the ordered (flat) state. Note that at $T=0$, a surface
described by a quantum SOS model always has to be in the ordered (flat)
phase \cite{IK} with a well-defined value of $\langle h \rangle$ and
saturation of correlation function $g_{jk}$ at large distances.

However, for $\beta \Gamma \ll 1$ the suppression of factor $K$ has to be
small, whereas the phase transition to the ordered state will take place
when this factor is suppressed from \makebox{$K_0=18$} down to
\makebox{$K^c_{1}=4$} (see Ref. \onlinecite{NHB}). Note that this has to
be so independently of whether in the ordered phase the spins on all three
sublattices are magnetically ordered or if on one of them the average
magnetization is zero (see discussion in Ref. \onlinecite{IM} and Ref.
\onlinecite{JE}). In terms of the SOS representation the first case
corresponds to having $\langle h\rangle$ integer and the second one to
having $\langle h\rangle$ half-integer, but in both cases the periodicity
of the $h$-dependent effective potential is the same, and the critical
value of $K$ is determined by this periodicity. \cite{Coleman}

Apparently such a pronounced suppression of $K$ cannot happen while
$\beta\Gamma\ll 1$ and requires $T\sim \Gamma$, whereas at $T\gg \Gamma$
the system has to remain in the critical phase. One can conclude that at
$J =\infty$ the phase transition between the critical and the ordered
states of the original spin model takes place at a finite temperature
$T^c_1$ proportional to $\Gamma$ whose value follows from the relation
$K(T^c_1)=K^c_1$.

\subsection{The case $J<\infty$}

When $J<\infty$, the Hilbert space of the model is substantially extended
because now all configurations of the Ising spins $\sigma_j=\pm 1$  are
allowed. In terms of the SOS representation, this corresponds to the
appearance of the possibility of the creation of screw dislocations
\cite{NHB} on going around which the integer variable $h_j$ interpreted as
height instead of returning to the same value changes by so-called Burgers
number $b=\pm 6$. Each dislocation is centered around a plaquette
containing not one but three \cite{comm-odd} frustrated bonds with
$\sigma_j\sigma_{j'}=+1$ which therefore can be identified with the
dislocation core. In the framework of the path integral description of a
quantum system, dislocations are linear topological excitations and have
either to form closed loops in space-time or to cross the whole system in
the direction of the Eucledean time.

In terms of the original spin variables the closed dislocation loops in
space-time correspond to the spin flips which are prohibited at
$J=\infty$. At $\Gamma\ll J$ such processes can be taken into account in
the framework of the perturbation theory. The most important of them is
the second-order process which leads to the decrease of the systems energy
by the amount proportional to $\Gamma^2/J$ per each spin which cannot be
flipped without increasing the energy of the system. At \makebox{$T\sim
\Gamma\ll J$} this gives a correction to $K$ of the order of $\Gamma/J$
and therefore leads only to a small shift of $T^c_1$ with respect to its
value at $J=\infty$.

On the other hand, the dislocations crossing the whole system in the time
direction can lead to the disordering of the critical phase. At
$\Gamma/J\to 0$, when the system does not experience evolution in the
Eucledean time these dislocations can be identified with that of the
classical SOS model. The core energy of such dislocations is proportional
to $J$, which makes their fugacities at $T\ll J$ exponentially small. The
main difference which appears at $\Gamma/J \ll 1$ is that quantum
fluctuations lead to a small negative correction to the core energy but
this is irrelevant for further reasoning.

When logarithmic interaction of dislocations is strong enough, they are
bound in neutral pairs and the system has the same properties as in the
absence of dislocations. \cite{NHB} The strength of this interaction is
determined by the same parameter $K$ as the amplitude of the fluctuations
of $h$. The dislocations with Burgers number $b$ remain bound in neutral
pairs only for
\begin{equation}                                         \label{Kc2}
K< K^c_{2} = \frac{b^2}{4}
\end{equation}
whereas at $K>K^c_2$ there appear free dislocations whose proliferation
leads to the disordering of the critical phase.  In the antiferromagnetic
Ising model on a triangular lattice dislocations have $b=\pm 6$ and,
accordingly, $K^c_{2}=9$ (as found in Ref. \onlinecite{NHB}).

Since $K^c_2=9$ is smaller than  $K_0 =18$ but larger than $K^c_1=4$, one
can conclude that at $\Gamma\ll J$, the dissociation of dislocation pairs
leading to disordering of the critical phase takes place at temperature
\makebox{$T^c_2\sim\Gamma$} which is higher than $T^c_1$ and therefore
there exists a finite interval of temperatures $T^c_1<T<T^c_2$ where the
system remains critical. The value of $T^c_2$ is determined by the
relation \makebox{$K(T^c_2)=K^c_2=9$}. As $T^c_1$, the transition
temperature $T^c_2$ is basically proportional to $\Gamma$ and remains
finite when $J$ is taken to infinity at a finite $\Gamma$ with the
difference between the values of $T^c_2$ at $J=\infty$ and at $\Gamma\ll
J<\infty$ being exponentially small in $J/\Gamma$. Note that there are no
reasons to expect the ratio $T^c_2/T^c_1$ to be equal to
\makebox{$K^c_2/K^c_1=9/4$} (as it was proposed in Ref. \onlinecite{JE})
because factor $K$ is not proportional to $T$ but depends on the ratio
$\Gamma/T$ in a more complicated way.

\section{Fully frustrated model on a square lattice}

The same approach can be applied to the fully frustrated transverse-field
Ising model on a square lattice. In the case $J=\infty$, the Hilbert space
of any fully frustrated transverse-field Ising model is defined by the set
of the ground states of the classical fully frustrated Ising model on the
same lattice, which is isomorphic to the full set of states of the
classical dimer model on the dual lattice. \cite{V77} On a square lattice,
the classical dimer model without any interaction of dimers is exactly
solvable \cite{F61} and allows for a mapping onto a SOS model \cite{H97}
with integer height variables $h_j$ defined on the sites of the same
square lattice. For a given configuration of Ising spins $\sigma_j$, the
values of $h_j$ can be defined by the relation analogous to Eq.
(\ref{h-h'}),
\begin{equation}                                               \label{h-h"}
    h_{j'}-h_j = -\left(1+2\tau_{jj'}\sigma_j\sigma_{j'}\right)=\left\{
    \begin{array}{l} +1\\-3\end{array}\right.
\end{equation}
where $\tau_{jj'}=J_{jj'}/J=\pm 1$. Eq. (\ref{h-h"}) assumes that the
square lattice is divided into two equivalent square sublattices (A and B)
and site $j$ belongs to sublattice A and site $j'$ to sublattice B. In the
opposite case the sign of $ h_{j'}-h_j$ would be the opposite.

When each square plaquette contains only one frustrated bond (with
$J_{jj'}\sigma_j\sigma_{j'}= J > 0$), Eqs. (\ref{h-h"}) unambiguously
define the values of all integer variables $h_j$ as soon as the value of
$h_j$ is chosen for one of the sites. On the other hand, any state of the
SOS model on a square lattice with $h_{j'}-h_j=+1,-3$ corresponds to some
ground state of the fully frustrated Ising model on the same lattice.

Since the classical system of noninteracting dimers on a square lattice is
in the critical phase, \cite{FS} the corresponding SOS model is in the
rough phase with the asymptotic behavior of the height-height correlation
function given by Eq. (\ref{g}), where according to Ref. \onlinecite{H97}
\makebox{$K=K_0=32$}. Like in the case of a triangular lattice, the
description in terms of the classical SOS model in which all allowed
configurations enter the partition with the same weight is applicable to
the fully frustrated transverse-field Ising model with $J=\infty$ in the
classical limit $\beta=0$ (that is, $T=\infty$).

Having  $T<\infty$ decreases the fluctuations of the surface. In the
lowest order in $\beta\Gamma$ the Hamiltonian of the classical SOS model
describing the system is again given by Eq. (\ref{Heff}). In terms of the
dimer representation, this Hamiltonian corresponds to having an attraction
between parallel dimers belonging to the same plaquette. The classical
dimer model with such an interaction has been studied in Ref.
\onlinecite{Alet}. In our system, the effective Hamiltonian has this form
only at the highest temperatures, whereas with the decrease of temperature
the interaction of the more distant dimers also starts to play a role.
However, since variables $h_j$ are integer, the phase transition to the
ordered phase takes place at the temperature at which factor $K$ is
decreased down to $K^c_1=4$ (exactly like in the case of a triangular
lattice), independently of what is the exact form of the dimer-dimer
interaction, In accordance with that, at $J=\infty$ the temperature of the
phase transition to the ordered phase is proportional to $\Gamma$ and for
$\Gamma \ll J<\infty$ the correction to the value of this quantity is
small.

The second phase transition is related to the appearance of free
(unpaired) dislocations crossing the whole system in the direction of
Euclidean time. Like in the case of a triangular lattice, dislocations can
be associated with the plaquettes which contain not one but three
frustrated bonds. \cite{comm-odd} For $K<K^c_2$ [where $K^c_2$ is given by
Eq. (\ref{Kc2})] they are bound in neutral pairs which dissociate when the
value of $K$ reaches $K^c_2$. In the SOS model describing the ground
states of the fully frustrated Ising model on a square lattice Burgers
numbers $b$ are equal to $\pm 8$ and, accordingly, $K^c_2=16$ (see Ref.
\onlinecite{Alet}). Since $K^c_2=16$ is smaller than $K_0=32$ but larger
than $K^c_1=4$, one can again make a conclusion that the critical phase
exists in a finite interval of temperatures $T^c_1<T<T^c_2$, whereas the
phase transition leading to the disordering of the critical  phase takes
place at temperature $T^c_2$ at which $K(T^c_2)=K^c_2=16$. For $\Gamma\ll
J$ this temperature like $T^c_1$ is proportional to $\Gamma$.

\vspace*{4mm}

\section{Conclusion}

In the present work we have demonstrated that in the quantum
antiferromagnetic transverse-field Ising model on a triangular lattice, as
well as in the fully frustrated Ising model on a square lattice, the
temperatures of the phase transitions separating the critical phase from
the ordered and disordered phases in weak fields are proportional to the
magnitude of the field. The analysis of Jiang and Emig \cite{JE} leading
to a different conclusion fails in weak fields ($\Gamma\ll J$) because the
continuous approximation used in Ref. \onlinecite{JE} for the description
of the spin fluctuations in Eucledean time requires the size of the system
in the time direction $\beta$ to be much larger than the typical time
between spin flips which is inversely proportional to $\Gamma$.
For $T\sim T^c_{1,2}$, this condition is fulfilled only when
$\Gamma_c-\Gamma\ll \Gamma_c$, whereas out of this range the continuous
approach cannot be trusted. The results of the numerical simulations of
the antiferromagnetic model on a triangular lattice  \cite{IM} are
consistent with $T^c_1$ and $T^c_2$ being proportional to $\Gamma$ at
small $\Gamma$.

$~$

The author is grateful to F. Mila and S. Wenzel for a useful discussion.


\begin{thebibliography}{9}
\bibitem{comm-odd} In the fully frustrated Ising models the number of
                   such bonds on each plaquette always has to be odd.
\end{thebibliography}
\end{document}